\documentclass[twocolumn,showpacs,preprintnumbers,amsmath,amssymb]{revtex4}
\usepackage{tabularx,graphicx}\begin{document}
\newcommand{\beq}{\begin{equation}}
\newcommand{\eeq}{\end{equation}}
\newcommand{\beqn}{\begin{eqnarray}}
\newcommand{\eeqn}{\end{eqnarray}}
\newcommand{\bmath}{\begin{subequations}}
\newcommand{\emath}{\end{subequations}}

\title{BCS theory of superconductivity: the world's largest Madoff scheme?}
\author{J. E. Hirsch}
\address{Department of Physics, University of California, San Diego\\
La Jolla, CA 92093-0319}

\begin{abstract} 
The time-tested BCS theory of superconductivity is generally accepted to be the correct theory of conventional superconductivity by physicists and, by extension, by the world at large.  
In a different realm of human activity, until very recently Bernard Madoff's time-tested investment operation was generally accepted as true and legitimate in the financial world. 
Madoff's  Ponzi scheme, where old investors were being paid off by funds contributed by
 new investors, 
was fundamentally flawed, yet was able to  thrive for decades because of many vested interests. `Red flags' suggesting its illegitimacy were ignored.
Here I suggest  that the same is true of  BCS theory.  There are an increasing number of  `red flags' that strongly suggest the possibility that BCS theory may be fundamentally flawed.
For example, an ever-growing number of superconductors are being classified as `unconventional', not
  described by the conventional BCS theory and each requiring a different physical mechanism. In addition, I argue that BCS theory is unable to explain the Meissner effect, $the$ most fundamental property of superconductors. There are several other phenomena in superconductors for which
 BCS theory provides no explanation, and BCS theory has proven unable to predict any new superconducting compounds.  
 From one day to the next, Madoff's edifice came crashing down and 
 a staggering 50 billion dollars evaporated, and I suggest that this may also be the fate of
 BCS theory. 
  I outline an alternative theory to conventional BCS theory  proposed to apply to all superconductors, `conventional' as well as `unconventional',  that offers an explanation for the Meissner effect as well as
  for other puzzles and provides clear guidelines in the search for new high temperature superconductors.
   \end{abstract}
\maketitle 

\section{Introduction}
In the progress of science, it is often the case that new theories supersede older theories without negating them.
Examples are  quantum mechanics and special relativity, which  
extended the range of validity of classical mechanics without negating  its validity  for  length scales and speeds familiar in everyday life. 
Then there are other cases where new theories negate older theories previously thought to be correct,
and replace them\cite{kuhn}. Examples of the latter
are Copernicus' theory negating Ptolemy's theory of planetary motion, Boyle's theory of caloric energy negating the phlogiston theory,  and Wegener's theory of continental drift
negating the theory of fixed continents with land bridges. There are many other such examples\cite{kuhn}.

Yet most working scientists appear to think that the first mode of scientific advancement is far more likely than the second.
In physics in particular, the great advances in modern physics in the 20th century occurred by superseding rather than negating previous theories,   and as a consequence physicists are especially disinclined to believe that contemporary  established theories could be
completely overhauled.
Evidence for this assertion is: attitudes of referees, journal editors, grant-allocating officers, conference organizers, as well as the research activities physicists choose to engage in indicate that they consider the possibility of an `established' scientific theory such as the BCS theory
of superconductivity  to be wrong to be nonexisting or of vanishingly small probability. 

Similarly, in the financial world, most investors until recently   believed that  large-scale  `Ponzi schemes' are a matter of the past (the original Ponzi scheme dates back to 1920).  That belief was shattered on December 11th, 2008, when it came to light that the thirty-plus-year-old  investment program run by highly respected stock market figure Bernard Madoff was nothing but an enormous Ponzi scheme, where older investors were being paid off with funds
collected from newer investors. On that day,
 fifty billion dollars suddenly evaporated, it was completely unexpected and has affected thousands including many `sophisticated' investors.

   Similarly I am suggesting here that tens of thousands of published papers, funding dollars and man-hours devoted to the BCS theory of 
 superconductivity over the past 50 years may evaporate from one day to the next  if BCS theory is
 proven wrong, either by an incontrovertible experiment or an alternative theory or both. 
 I argue that BCS theory has an unrecognized fundamental flaw, its inability to explain the most fundamental property of superconductors,
 the Meissner effect, and that this calls the validity of the entire framework into question, including the validity of London's electrodynamic description of superconductors\cite{london}.  
Furthermore, BCS theory is completely unable to predict superconductivity in new materials.
I discuss many other reasons that make the BCS scheme suspect, and 
 point out many similarities between the current status of BCS theory and the Madoff operation pre-December 11, 2008.

  Just like physicists today are absolutely convinced that BCS theory is correct, Madoff's investors were also absolutely convinced  yesterday that Madoff's scheme was `correct', otherwise they would not have entrusted their money to him, in some cases their entire wealth. 
 They did not have ready access to information that existed that could have suggested otherwise, and the vast majority of them didn't know there was any reason to 
 spend time or effort looking for such information.

  The possibility that Madoff was a fraud was, however, forcefully suggested by H. Markopolos in 1999 in a communication to the Securities and Exchange Commission (SEC), and he continued gathering evidence in support of his
 contention and attempting to prompt action against Madoff for several years, culminating in 2005 with his memo to the SEC 
 entitled ``The world's largest hedge fund
                  is a fraud''\cite{markopolos}, where he listed 29 `red flags' to support his contention. But,  Markopolos'  calls for actions
 went unheeded for nine years, during which investors continued to pour billions of
 dollars into the scheme, until it suddenly collapsed  on its own.
 Today, financial publications are full of analyses of red flags that were overlooked.
Similarly in science, as argued by Lightman and Gingerich\cite{anomalies}, anomalies (red flags) are often widely recognized as such only $after$
a new theoretical framework is found that explains them. They coined the term `retrorecognition' for this phenomenon.

  Just as Markopolos who was prompted to action by his inability to reproduce Madoff's purported success in his own investment activities, I started to doubt the validity of BCS theory
  many years ago
  when I was unable to reproduce superconducting behavior in my research work involving
  numerical simulations. Over the past 20 years, I have invested my scientific activity
  in proposing an alternative theory of superconductivity which is incompatible with conventional BCS theory\cite{mytheory}, hence I am certainly not an `unbiased' observer. 
  Nevertheless  I hope readers will spend some time considering the points raised in this paper and
 do followup checking on their own. 
 
In this paper what I mean by `BCS theory' is   the  
BCS pairing theory through the electron-phonon interaction mechanism as formulated in the original BCS paper\cite{bcs}, and its extension
to include the effect of a retarded interaction, generally known as Migdal-Eliashberg theory\cite{migdal,eliash}. This theoretical framework is generally believed to describe
the superconductivity of `conventional' superconductors, both type I and type II, including all the elements and thousands of 
compounds\cite{allenmitrovic,carbotte0}.   
Then there are other classes of materials discovered in recent years generally believed not to be described by BCS theory, as discussed later in this paper.

 There is of course a fundamental difference between both situations. Madoff was the central figure  that deliberately misled investors, and  I am
 certainly  not suggesting that there is  an analogous figure in BCS theory:
 there isn't.
 But I am suggesting that many participants unwittingly or  perhaps in some cases half-wittingly aided, and thus enabled, the deception in both situations  for the same self-serving reasons.
 The purpose of comparing the Madoff and BCS situations is: 
having now the benefit of hindsight in interpreting warning signals in the Madoff situation that could have prevented the scheme from continuing for so long had they been
 correctly interpreted, to the extent that physicists recognize that similar factors may be at play in the   BCS situation  would encourage
 a serious examination of many warning signals that I assert exist in the BCS situation but are being ignored.

        \setcounter{secnumdepth}{1}

 \section{Why people believed in Madoff and why people believe in BCS theory}

  There are of course good reasons why a set of incorrect beliefs can go unchallenged for a long time\cite{shermer}.
 Here I list some of the factors that I submit have made BCS theory successful for so long without being necessarily correct,
 and analogous factors at play in the Madoff scheme.
   
 \subsection{1. Kernel of truth}
Even in the original Ponzi scheme there was a kernel of truth: In 1920, international reply coupons (IRC) entitled mail recipients to use them as postage of a reply; the differential pricing of IRC's in 
 different countries allowed for a potential profit, so Ponzi bought IRC's at a low price in Italy and exchanged them for higher value US stamps. 
 Many  such `arbitrage' strategies allowing players to profit  from inefficiencies still exist in financial markets today.  Madoff's 
 `split-strike'  options strategy presumably delivered steady positive results over some period of time.   
 
Similarly, parts of BCS theory are certainly correct and represented an important advance when first proposed: the concepts of Cooper pairs, of macroscopic phase coherence,
 and the existence of an energy gap are incontrovertible.  These elements of the theory led to explanation and even prediction of puzzling experimental data
 such as NMR relaxation rate\cite{nmr} and Josephson tunneling\cite{josephson}.
 However many other aspects of BCS theory and especially the electron-phonon mechanism 
 I contend are not correct despite being universally accepted.
 
 The fact that part of a scheme is believable does not make the entire scheme believable. 
 By the end of Ponzi's scheme, the number of IRC's that would have been needed
 to be circulating was 6,000 times larger than were actually in circulation. In Madoff's case, the size of his operation already in 2005 would have required
 many more call options in the stock exchange than were actually outstanding\cite{markopolos}.
Similarly, the BCS electron-phonon mechanism of superconductivity may have been convincing
 around 1970 as a `universal' mechanism for all known superconductors\cite{parks}. By now, as discussed below, 
 there are at least ten different classes of materials that clearly cannot be explained by the electron-phonon mechanism, each 
 requiring its own different mechanism if BCS theory is assumed to be correct.

  \subsection{2. Respectability of key actor}
    Bernard Madoff was beyond suspicion because he was highly respected in the securities industry due to  his long and distinguished career: he had played a leading role in the development of the   Nasdaq stock market, served as its Chairman, pioneered electronic trading and owned one of the largest market-maker firms on Wall Street.
    
Similarly John Bardeen was `beyond suspicion': just the year before he proposed BCS theory (1957), John Bardeen had been  awarded the Nobel prize in physics
for the invention of the transistor;  he had  had a long and 
  distinguished career in theoretical physics, and had been working and publishing on the problem of superconductivity for over twenty years. 
In 1956   he had published an authoritative review on superconductivity\cite{bardeenreview}. The fact that Bardeen was regarded as an authority in superconductivity at the time is evidenced by the fact that the New York Times wrote a story 
  on the BCS theory of superconductivity less than a month after it appeared in print\cite{plumb}.
  
    \subsection{3. Early doubters proven wrong}
        As early as 1992 Madoff was investigated by the SEC, accused of dealing with unregistered securities and suspected of running a Ponzi scheme. However, he showed that he had indeed delivered the returns
    that clients had been promised, and was cleared of any wrongdoing. It is likely that the negative result of that inquiry discouraged the SEC from examining Madoff again later.
    
Similarly, there were early doubts about the validity of BCS theory because its `proof' of the Meissner effect
    failed to satisfy gauge invariance\cite{gaugedoubts}. However, it was later shown that
    the BCS derivation was valid in the particular case of a transverse gauge and plausible arguments were given for generalizing the theory to
    an arbitrary gauge\cite{gaugenodoubts}. Thus the early doubts were allayed and as a consequence  the theory became more firmly established.
    
As I will argue later, these early discussions did not really address the essence of the Meissner effect, which remained unexplained
    within BCS theory. But the fact that the early doubts had been resolved undoubtedly led to the general belief that  all doubts concerning
    the Meissner effect within BCS had been discussed at length and resolved and there was no point to rehash them.

    \subsection{4. Selected few get to participate}
        One of the key attractions of the Madoff investment scheme was that it was hard to `get in'. Madoff reportedly turned down many investors who  wanted to
    invest with him, and thus those that were accepted  felt privileged for being `in' and were discouraged from asking questions for fear of being `kicked out'.
    
Similarly, one doesn't become an expert in BCS theory overnight. One needs a background in many-body theory and second quantization as well as
in solid state physics and statistical physics. Concepts such as off-diagonal long range order and broken gauge 
invariance are rather subtle. Beginning students asking  interesting questions such as how can  one possibly explain the Meissner effect, or why the theory is unable to predict
new superconductors, are told to wait until they master the advanced  mathematics and physics
 required to really understand it, or else go elsewhere. By the time they have mastered this technology they have
forgotten the interesting questions they had or have convinced themselves that they are no longer relevant.

      \subsection{5. Gatekeepers and non-gate-keeper participants}
            The vast majority of Madoff investors did not know much about Madoff. They invested in Madoff's operation through `feeder funds'  and trusted
      the managers of the feeder funds as well as their well-known auditing firms.
      There is no allegation that these `gatekeepers' were privy to Madoff's deception. However, it is clear that they greatly benefitted from the arrangement by
      collecting huge manager fees with very little work, hence they had a huge material disincentive to raise any questions about Madoff .      
      The investors in turn trusted the expertise of the gatekeepers and being less expert in financial matters than the feeder  fund managers 
      saw no reason to spend their time evaluating the trustworthiness of Madoff 's operation themselves.
 
      In the BCS case, the `gatekeepers' are those relatively few who  have themselves performed detailed Eliashberg calculations of  first-principles  bandstructures 
      and electron-phonon interaction parameters to calculate  superconducting properties of real materials.  
      The vast majority of physicists that use BCS theory do so with model Hamiltonians that don't have a clearcut justification nor very direct connection to real materials.        
      The gatekeepers tell us that their calculations reproduce the measured superconducting $T_c$'s, gaps, isotope effect, structure in tunneling characteristics, etc. of real materials,
      and thus prove beyond doubt that BCS-electron-phonon  theory describes conventional superconductors.
      The rest of physicists   blindly trust the gatekeeper's statements.
            
However,  just like in the case of Madoff, the BCS `gatekeepers' have a lot to lose from BCS theory being wrong. They have invested  considerable time and effort in 
      becoming expert in these calculations, and benefit from the status quo. They have funding to perform such work, their work is being cited by the non-gate-keeper participants,
      and their careers advance.  They are the best qualified to question  BCS theory but have no incentive to do so.

                  \subsection{6. Red flags and early whistleblowers}
             As mentioned earlier, Markopolos became convinced as early as 1999 that the Madoff scheme was a scam. He contacted the SEC
                  repeatedly between 1999 and 2008, to no avail. In 2005 he sent a 19-page complaint to the SEC entitled ``The world's largest hedge fund
                  is a fraud'', detailing 29 `red flags' in support of his contention\cite{markopolos}. 
                  
                  Questions were also raised by others. The respected financial publication Barrons wrote an article about Madoff in May 2001 where it
                  suggested that Madoff was subsidizing his investment operation with his market-making activity, but did not raise the possibility
                  of fraud. The same month, Michael Ocrant wrote an article in Managed Account Reports entitled ``Madoff tops charts; skeptics ask
                  how''  suggesting a similar explanation for Madoff's amazing performance,  
                  also not raising the possibility of fraud.

 In the case of BCS, the theory  was widely accepted soon after publication but some early questions were raised whether the
                  electron-phonon mechanism applied to the transition metal superconductors\cite{kondo,suhl,jensen}. However,  by 1969
                  when Park's treatise on superconductivity
                  was published\cite{parks} it was universally accepted that BCS-electron-phonon theory described all known superconductors. 
                  
                  Except for
   one  persistent gadfly: Bernd Matthias, a well-respected solid state experimentalist who had been 
                  making superconducting materials in his lab  for many years\cite{matthias}.
                  In paper after paper 
                  and conference proceedings after conference proceedings  in the 60's and 70's Matthias argued that BCS theory could not
                  possibly be the correct theory of superconductivity because it was unable to predict   new superconducting materials. 
                  Matthias found many new superconductors through  empirical rules that he deviced, but found no guidance whatsoever in BCS theory.
                  The physics community politely tolerated Matthias' rantings and ravings but he did not produce any followers. 
                  When he passed away in 1980, the sole voice calling into question BCS theory went silent.
                  
                  In 1988 I came to the conclusion that BCS theory is incorrect, shortly after the discovery of the high temperature superconductors by
                  Bednorz and Muller in 1986\cite{holesc1}. I have written many papers since then developing a new theory and pointing out many  anomalies in BCS theory.
                  Nevertheless BCS theory remains as firmly established today as Madoff's investment scheme was on December 10th, 2008, 
              nine years after Markopolos had   identified it as a fraud and one day
                  before its demise.

              \subsection{7. Role of mainstream media}
              
              There was no follow-up to the Barrons'  2001 story by Barrons in later years nor anywhere else in the mainstream media. Given
              the magnitude of Madoff's investment operation this is
              remarkable.  Similarly in the case of BCS, the `mainstream media', meaning the most prestigious physics publications such as Physical Review Letters,
              Science, Nature, PNAS, Physical Review B, etc, are silent about the possibility that BCS theory could be wrong, while being
              full of papers devoted to applications of BCS theory.      Papers submitted to these journals casting doubt on the validity of
              BCS theory to explain conventional superconductors are rejected\cite{silence}.

                \subsection{8. The `proof' of the validity of a flawed scheme}
                There is no shortage of `proofs' of flawed schemes before their invalidity is discovered. 
                In the Aristotelian-Ptolemic geocentric theory, `proof' that the earth was at rest was the absence of `wind' and the apparent absence of motion of the fixed stars.
               In the  book ``Why People Believe Weird Things: Pseudoscience, Superstition, and Other Confusions of Our Time'' the author gives many examples
               of flawed `proofs' of invalid beliefs\cite{shermer}.
                
  The most quoted reason given as convincing proof that BCS-electron-phonon theory describes conventional superconductors is the structure in tunneling characteristics detected
  in normal-insulator-superconductor tunneling experiments, where small wiggles in the tunneling conductance as function of voltage match the peaks and valleys of 
  the phonon density of states as function of frequency measured in neutron scattering experiments in several materials, most notably Pb\cite{tunnelingpb,scalapino,mcmillanrowell}.
  
  I am not disputing the interpretation that the structure in the tunneling conductance reflects the phonon spectrum. As Bernd Matthias said\cite{matthias},
  ``you can't ever stop a crystal lattice from rattling''. Even the gap of ordinary semiconductors is modulated (but not caused!) by the electron-phonon interaction and
  shows an isotope effect\cite{haller}. What I am disputing is the interpretation that the small modulation (few $\%$) of the tunneling conductance spectrum by the phonons
  is $proof$ that superconductivity is caused by lattice vibrations and would not exist for infinite ionic mass.
  
  The interpretation of tunneling results is cast in terms of the spectral function $\alpha^2 F(\omega$), where $F(\omega)$ is the phonon spectral function determined from
  neutron scattering experiments. What is $not$ emphasized is that $\alpha^2$ is itself often a strong function of $\omega$ that is not directly accessible
  to experiment\cite{weber}.
                    
                \subsection{9. Long timescale}
   One of the arguments physicists would give to deny the possibility that BCS theory could be wrong is that it has been around for
                so long, over 50 years. Similarly, before Madoff's scheme imploded, financial experts would have said that a Ponzi scheme
                cannot possibly go on for 30 years. Now we know better. Because of the large number of vested interests and 
                highly motivated gatekeepers that
                develop around such a scheme in our modern world, the timescale for uncovering such a financial scheme or for debunking
                an established scientific theory that is incorrect, may have become longer than 
                anyone would have expected.
                                
                  \subsection{10. BCS theory as a `Ponzi scheme'}
                  
                 In a financial `Ponzi scheme', old investors are paid off by funds contributed by new investors. The old investors spread
                  the word that this is a good scheme and this induces more new investors to come in.  
                  I am certainly $not$  suggesting that there is deliberate deception in the case of a scientific theory such as BCS, still
                  I argue that a similar phenomenon occurs.                  
                  The payoff to the old `investors' (established physicists) 
                  comes in the form of citations to their papers by younger physicists and awards of grant money  through which the older physicists
                  are expected to train the new generation of physicists. The grant money   also provides for Summer salary, equipment, travel funds  and other perks                  for the older physicists.
                  These payoffs depend on the existence of a crowd of younger physicists eager
                  to get into the game and continue building up the theory, 
                  lured by the success of the older physicists as evidenced by their career advancement, prestige, prizes, etc.
                  Questioning of the old theory is discouraged in many ways, and   early questioning would result in the young
                  physicist being  denied career opportunities open to his/her non-questioning peers.                  The flawed  scheme continues building up and reinforced by those that
                  are allowed to enter, and everybody turns a blind
                  eye to anomalies that could suggest something is wrong\cite{anomalies}. There are however many such anomalies (red flags) in the case of 
                  BCS theory, as detailed in the next section.

  \section{Red flags in BCS theory}
   Markopolos pointed out  29 red flags   in the Madoff case\cite{markopolos}. I point out the following 10 in the BCS case: 
     \subsection{1. Lack of transparency}
 One important red flag in the Madoff case was lack of transparency. Madoff refused to disclose any details of his investment scheme,
  other than it was based on a `split-strike' options strategy, and never reported what
  investment positions he took. Prospective investors asking for a more detailed explanation of the investment strategy were told
  they could not invest.

  It can also be said about BCS theory that it is anything but transparent. It is extremely hard to explain it  to a non-physicist and even to a
  non-solid-state physicist, and it defies intuition. How can the very strong direct Coulomb repulsion between electrons
  be overcome by a small `second-order' electron-ion induced attraction? Why are some materials not superconducting at any temperature?  How is it that sometimes a high phonon frequency leads to 
  high $T_c$\cite{highfreq,metallich} and sometimes a low phonon frequency (the soft-phonon story\cite{softph}) leads to high $T_c$? 
  
  There is no simple intuitive criterion in BCS theory that allows to understand qualitative trends in $T_c$ in materials. The
  Debye-frequency prefactor in the BCS expression for the critical temperature suggests that going down a column in the periodic table 
  (where elements have the same valence-electron configuration) $T_c$ should decrease due to the increasing ionic mass. 
  This is $not$ what happens\cite{debye}. There are no qualitative criteria that can be used to estimate even the order of magnitude of critical temperatures, nor whether a material
  is or is not a superconductor.
  The gatekeeper `experts' tell us that $T_c$'s depend on many subtle details and can go up and down with different combinations of phonon frequencies,
  electron-phonon coupling constants, band structure details, strength of Coulomb interactions and of spin fluctuations, 
  etc\cite{allenmitrovic,scalapino,mcmillan,carbotte,allendynes,papa32,papahcp}. 
  The `Coulomb pseudopotential' serves as the wildcard that ensures that theory will always fit experiment\cite{li,pseudopot}.
  
   \subsection{2. Increasing number of epicycles}
   Given that initially the isotope effect was claimed to be the `proof' that the electron-phonon interaction is responsible for
  superconductivity, 
 an early observation not easily explained by BCS theory was the absence of isotope effect in certain elements like ruthenium\cite{ru} and 
  osmium\cite{osmium}
  and an inverse isotope effect in uranium\cite{uranium}. However, it was argued that more elaborate versions of the theory could account for the 
  observations\cite{garland,uraniumexplained}.

  Another observation calling into doubt the conventional theory was the absence of a strong electron-phonon structure in the tunneling spectra
  of niobium\cite{bostock,bostock2}, the element with the highest $T_c$. However it was argued that a more elaborate theory taking into account the proximity effect due to the
  complicated nature of the tunnel junctions could explain the observations\cite{wolf}. 
  
  The early transition metals $Sc$ and $Y$ as well as the late transition metals like $Pd$ are not superconducting at ambient pressure, 
  even though they would be expected to be so given their other properties, according to the conventional theory\cite{parks2}.
  To explain this, it is necessary to invoke the Coulomb pseudopotential `wild card', and it
 is argued that `antiferromagnetic spin fluctuations' suppress the expected superconductivity of scandium and yttrium\cite{capellmann}, 
  and `ferromagnetic spin fluctuations' suppress the expected superconductivity of palladium\cite{pd}. 
However it  is not explained  why these fluctuations do not give rise to `unconventional' superconductivity in those elements. For example, it was suggested for $Pd$ a propensity to p-wave superconductivity induced by
  ferromagnetic spin fluctuations\cite{fayappel}. This was however disproved by the finding of s-wave 
  superconductivity in $irradiated$ $Pd$ at $3.2K$\cite{irradiatedpd}. Furthermore, some of those elements were recently found to display quite high superconducting transition temperatures
  under pressure (not predicted by theory), as discussed in the next section.

  In 1969 when Parks' treatise on superconductivity was published\cite{parks}, there was general agreement that BCS theory with the electron-phonon
  mechanism explained all known superconductors. Particularly interesting is the article in that treatise by Gladstone, Jensen and Schrieffer
  on ``Superconductivity in the Transition Metals''\cite{parks2}. As  mentioned earlier,   doubts had been raised by Bernd Matthias and others whether other
  mechanisms of pairing may be at play in transition metals\cite{matthias,kondo,suhl,jensen}, which were reviewed in this article
  and dismissed. In fact one of
  its authors,  Jensen, had been one of the early questioners of BCS-electron-phonon mechanism 
   for Lanthanum and Uranium\cite{jensen}.
  However by 1969 he clearly had been brought `into the fold': the Gladstone et al paper concludes, referring to predictions of
  non-electron-phonon superconductivity in Lanthanum,  
  ``Although initially these predictions appeared to be found experimentally, more recent work on cleaner samples gives no evidence
  that La is anything but a phonon-induced BCS superconductor'', and similarly for all other transition metals.

  However, since 1970   at least 10 distinct materials or families of materials have been discovered that exhibit superconductivity
  for which there is a consensus that they cannot be   described by the electron-phonon BCS theory, 
  or at least there are serious doubts whether they can, namely: 
  (1) High $T_c$ cuprates, hole-doped ($YBa_2Cu_3O_{7}$) and electron-doped ($Nd_{1-x}Ce0_xCuO_{4-y}$); (2) Heavy fermion materials ($CeCu_2Si_2$, $UBe_{13}$, $U Pt_3$); (3) Organics ($TMTSF_2PF_6$); 
  (4) Strontium-ruthenate ($Sr_2RuO_4$); (5) Fullerenes ($K_3C_{60}$, $Cs_3C_{60}$); 
  (6) Borocarbides ($LuNi_2B_2C$, $Y Pd_2 B_2 C$); (7) Bismuthates ($Ba_{1-x}K_x BiO_3, BaPb_{1-x} Bi_x O_3$);
  (8) 'Almost' heavy fermions ($U_6Fe$, $U Ru_2 Si_2$, $U Pd_2 Al_3$);
  (9) Iron arsenide compounds ($LaFeAsO_{1-x}F_x$, $La_{1-x}Sr_x  Fe As$);  (10) Ferromagnetic superconductors ($UGe_2$, $U RhGe_2$).
  In addition, magnesium diboride ($MgB_2$) was believed initially to be outside the scope of BCS electron-phonon theory, however that has changed by now. We return to
  this interesting material in the next subsection.
  
  The ten materials or classes of materials listed above
  exhibit each different deviations from conventional BCS behavior, and/or their $T_c$ is too high to be described by BCS-electron-phonon theory,
  however   there is also no indication that they can all be described by
a single alternative mechanism or theory. Rather, new different mechanisms and theories have been proposed to describe each of these situations. 
  If BCS theory is correct for the conventional superconductors, we would need new different theories to describe
  d-wave symmetry states, p-wave symmetry states, superconductivity arising near a Mott insulating state, 
  antiferromagnetic-spin-fluctuation induced superconductivity, ferromagnetic-spin-fluctuation induced superconductivity, 
  superconductivity induced by low dimensionality, charge-density-wave induced superconductivity,
  superconductivity induced by inhomogeneity (stripes), d-density waves, quantum critical points, 
  marginal Fermi liquids, superconductivity with and without `glue',
  resonating-valence-bond-induced superconductivity, etc. etc. to encompass all these new materials discovered since 1970.

 The Proceedings of the series conference ``Materials and Mechanisms of Superconductivity'', held every three years since 1988, and earlier the Proceedings of the d- and f-band superconductivity
 conferences held every two or three years since 1971, 
provide a large number of  references for these multiplying entities.
   
  The situation is analogous to the situation in astronomy shortly before the advent of Copernican theory.
  To explain an increasing number of astronomical observations
using the Ptolemy
  paradigm of the earth as the center of the universe prevalent at the time, increasingly more complicated models postulating
  an increasing number of epicycles to describe
  retrograde motion of planets had to be introduced. 
  Similarly, for each new observation unexpected within the conventional BCS theory
  a new twist is added to the theory to explain the observation, or else the material is declared to be `unconventional',
  hence not described by conventional BCS-electron-phonon theory. 
  The validity of conventional BCS theory for `conventional' superconductors is $never$ questioned.

  \subsection{3. Inability to predict yet ability to post-dict}
  
Matthias repeatedly emphasized that BCS theory and its implications did no lead to the ability to $predict$ whether
  a compound or a family of compounds would be superconducting. The situation has become even far more egregious 
since the 70's up to today, with the advent of an ever-increasing number of `unconventional' superconductors
and the discovery of substantially higher temperature superconductivity in `conventional' superconductors under applied high pressure.
  
  For a while, the search for new higher $T_c$ superconductors was directed at compounds with light elements, that would give rise to a high
  Debye frequency which appears as a prefactor in the BCS expression for $T_c$. High $T_c$ superconductivity was predicted for
  metallic hydrogen\cite{metallich} and for metal hydrides\cite{hydrides}. 
  Indeed, superconductivity around $10K$ was found in thorium-hydride\cite{pdh0} and in palladium-hydride\cite{pd0prime}.
  Of course it was very disappointing when substitution of hydrogen by the heavier isotope deuterium gave an even
  higher $T_c$\cite{pdh}, but theory found a ready way to explain it\cite{pdh2,papa}, and even to this day theorists continue `predicting' that metal hydrides
  will yield high temperature superconductors\cite{pdh3}.
  
 Similarly, superconductivity was predicted   for the light metal Lithium, the simplest of simple metals, at ambient pressure with critical temperature 1K or higher \cite{li,li2}. After many years,
 superconductivity at ambient pressure in Li was found but only at temperatures below 0.0004K\cite{lifound}.
 
  High $T_c$ was predicted in quasi-one-dimensional materials, based on Little's excitonic mechanism for superconductivity\cite{little}.
  None of it was found.
  
  Instead, a ``soft-phonon'' scenario was developed to `predict' relatively high $T_c's$ in materials with low frequency phonons\cite{mcmillan,soft2}, in response to the 
  experimental findings of such materials, e.g. the A15 family of compounds\cite{testardi}.
  
  In 1972, Marvin Cohen and Phil Anderson `predicted' that superconductivity with critical temperatures much above what existed at the time ($\sim 20K$)
  was impossible in any material\cite{cohenanderson}, through the electron-phonon or any other mechanism.  This did not prevent Time Magazine from reporting in 1987,
  shortly after superconductivity above $90K$ was experimentally discovered\cite{chu},  that
  {\it ``At the University of California, Berkeley, a group that included Theoretical Physicist Marvin Cohen, who had been among those predicting superconductivity in the oxides two decades ago, reproduced the 98 K record, then started trying to beat it.''}\cite{time}
However, the first paper written by  Cohen discussing superconductivity in an oxide  was in 1964\cite{oxides}, where he discussed the
  just discovered superconductivity with $T_c=0.28K$ in semiconducting $SrTiO_3$  and referred to his earlier work on possible superconductivity in
  semiconductors that $did$ $not$ mention $either$ semiconducting $or$ superconducting   oxides. Subsequently Cohen `predicted' the carrier concentration dependence
  of $T_c$ in $Sr_2RuO_4$, including its maximum at $\sim 0.30 K$, $after$ it had been experimentally measured\cite{oxides2}. Never did Cohen consider
 in his printed work the possibility of superconductivity in oxides at higher temperatures until after it was experimentally discovered.
  
  Magnesium diboride ($MgB_2$) was found to be superconducting in 2001 with a critical temperature of $39K$\cite{mgb20}, completely unprecedented for
  a metallic compound with only s- and p-electrons. It was not predicted by theory, and it exhibits a small isotope effect.
  Nevertheless this has not prevented theorists from claiming  that the conventional BCS-electron-phonon theory completely explains the 
  high $T_c$ of $MgB_2$\cite{mgb21,mgb22,mgb23,mgb24}. Based on these calculations theorists have now predicted higher $T_c$ superconductivity in
  related compounds such as $Li_{1-x}BC$\cite{predlibc,prediclibcprime,predlibc2} and in $BC_3$\cite{predbc3,predbc3prime}. None has been found in either 
  system\cite{cit1,cit2}.
  
  As mentioned in the previous section, Scandium is not superconducting at ambient pressure, and this is `explained' by the Coulomb pseudopotential
  wildcard\cite{capellmann}. In 1979, Sc under pressure ($\sim 200 kbar$) was found to be superconducting with
  $T_c\sim 0.35 K$\cite{wittig}, and in 2007, its critical temperature  was found to rise to $8.2K$ at pressures of $740 kbar$\cite{hamlin1}.
  None of this was predicted by theory, but subsequently calculated  and claimed  to be `in good agreement with experiment' \cite{papasc}.
However, shortly thereafter, Scandium's critical temperature rose by over a factor of 2, to  $19.6K$ at $1 Mbar$ pressure\cite{hamlin2}. Presumably we will see shortly
  a theoretical `prediction' of this remarkable increase. 
  
More generally, there have been remarkable advances in achieving superconductivity with higher transition temperatures in the elements under high pressure in recent years, e.g.\cite{hamlin2,pressure}:
lithium, $T_c=16K$  ($T_c=0$) at $800 kbar$ (at ambient pressure); boron,  $11K$ ($0$) at $250 kbar$; 
sulphur, $T_c=17.3K$ ($0$) at $1.9 Mbar$; calcium, $T_c=25K$ ($0$) at $1.6 Mbar$ ; yttrium, $T_c=19.5K$ ($0$) at $1.1 Mbar$; 
lutecium, $T_c=12.4K$ ($0$) at $1.7 Mbar$; vanadium, $T_c=16.5K$ (5.4K) at $1.2 Mbar$; 
zirconium,  $T_c=11K$ ($0.55K$) at $300 kbar$. None of these have been predicted by theory, but there is
an ever-increasing number of theoretical `post-dictions' of the 
observations\cite{pdb,pdli,pdli2,pdli3,pdli4,pdyca,pdy2,pdy3}.

For example, in a postdictive study of Yttrium under pressure, it is claimed that theoretical calculations
`provide a good interpretation of the measured increase of $T_c$  in these
metals'\cite{pdyca}, yet the results shown indicate that even an anomalously low Coulomb pseudopotential
$\mu ^*\sim 0.04$ yields a critical temperature substantially lower than the observed one\cite{pdyca}.
Another postdictive calculation for Y under pressure claims that it
`demonstrates
strong electron-phonon coupling in this system that can account for the observed range of $T_c$' using a Coulomb 
pseudopotential value $\mu ^*=0.15$\cite{pdy2}, while acknowledging that their
more detailed approach
`has not yet provided $-$ even for elemental
superconductors $-$ the physical picture and simple trends that
would enable us to claim that we have a clear understanding
of strong-coupling superconductivity'\cite{pdy2}.

  \subsection{4. Blind use of formalism}
  
  In order to explain the increasingly higher $T_c's$ found in supposedly `conventional' materials, higher values of the
  electron-phonon coupling constant $\lambda$ have to be used in the conventional
  formalism\cite{allendynes}. In fact, as early as 1975 values of $\lambda$ as high as $2.5$ were
  postulated to explain the $T_c$ of $Pb-Bi$ alloys\cite{allendynes}. To explain the superconductivity of $Y$ under pressure
  a value of $\lambda=2.8$ is used\cite{pdy2}, and $\lambda$ as high as $3.1$ is assumed to explain the
  superconductivity of $Li$ under pressure\cite{pdli3}. However, it
  has been convincingly shown analytically\cite{alexandrov} that $\lambda$ values larger than $\sim 1$ {\it should not be used} in the conventional 
  formalism, because for $\lambda>1$ the electron-ion system collapses to a narrow band of small polarons, whose description is
  outside the reach of the conventional theory. 
  This result is  confirmed by numerical simulation studies\cite{holstein}. This finding is completely ignored and the conventional formalism continues to be
  routinely used irrespective of whether $\lambda$ is small or large.    
  
                        \subsection{5. Inability to explain Chapnik's rule}
           There is a simple empirical rule that can predict with good accuracy whether or not a material is superconducting: the sign of its Hall coefficient.      
           The vast majority of superconductors have positive Hall coefficient in the normal state, indicating that the transport of current occurs through
           holes rather than electrons\cite{chapnik,chapnik2,chapnik3}. The electron-doped cuprate superconductors
           only become superconducting in the doping and reduction regime where their
           Hall coefficient changes sign from negative to positive\cite{electrondoped,electrondoped2}.
          The sign of the Hall coefficient is a far better predictor of whether a material is or is not a superconductor than any other normal state
           property\cite{correlations}, yet the conventional BCS-electron-phonon theory has no explanation for this observation.           It would be of great interest to measure the Hall coefficient of non-superconducting elements that become superconducting
           under applied pressure, which should give further evidence for this correlation between the
           character of the normal state charge carriers and superconductivity.
           
           \subsection{6. Inability to explain the Tao effect}
           
           In a series of experiments beginning in 1999, Rongjia Tao and co-workers found that millions of 
           superconducting microparticles in the presence of a strong electrostatic field aggregate into balls of macroscopic dimensions\cite{tao1,tao2,tao3}. No explanation of this phenomenon exists within the conventional theory of superconductivity.
           Initially the finding was attributed to special properties of high $T_c$  cuprates,
in particular, their layered structure\cite{tao1}, however,
subsequent experiments  for conventional
superconducting materials  all showed the same behavior\cite{tao2,tao3}.         
           
           The conventional
theory of superconductivity predicts that superconductors
respond to applied electrostatic fields in the same way as
normal metals do\cite{tao4,tao4prime}, by forming chainlike structures. Hence  Tao$'$s observation represents a fundamental puzzle within
the conventional understanding of superconductivity, yet no explanation of the  effect has been proposed by defenders of the
           conventional theory of superconductivity.
           The response of superconductors to applied  electric fields is as fundamental a question as their response to applied magnetic fields.

     \subsection{7. Inability to explain the De Heer  effect}
     
     In a series of experiments, De Heer and coworkers have discovered that small Niobium clusters at low temperatures develop ferroelectric dipole 
     moments\cite{deheer1,deheer2,deheer3}.  They find strong evidence that the electric dipole moment is
     associated with pairing of valence electrons and mirrors superconducting properties of the bulk material. Such behavior is
     unexpected both for a normal metal as well as for a superconductor, and suggest a fundamental inadequacy of the conventional theory of superconductivity.
     The same behavior is found by De Heer in alloy clusters of Nb and in clusters of other transition metals that are superconducting in the bulk.

                      \subsection{8. Inability to explain rotating superconductors}
                      
A superconducting body rotating  with  angular
velocity $\vec{\omega}$ develops a uniform magnetic field throughout its interior given by\cite{londonm,londonmprime}
\beq
\vec{B}=-\frac{2m_e c}{e}\vec{\omega}
\eeq
 where $e$ and $m_e$ are  the charge and mass of the superfluid charge carrier respectively,  and $c$ is  the speed of light.    
    This has been determined experimentally    for both conventional superconductors\cite{lm1,lm2,lm3},   heavy fermion\cite{lm4}  and high $T_c$\cite{lm5}  superconductors.
    The associated magnetic moment is termed the `London moment'.  
 
What is remarkable about this observation is: (i) The measured magnetic field is always $parallel$, never $antiparallel$ to the angular velocity. 
This implies that the superfluid charge carriers have negative charge, i.e. they are electrons, not holes.
This is despite the fact that the normal state carriers in all these materials are holes. (ii) The mass and the charge entering Eq. (1) correspond to the
{\it free electron mass and charge}.
(iii) The magnetic field Eq. (1)  is the same whether a superconductor is put into rotation or a rotating normal metal is cooled into the superconducting state.

 The fact that it is the electron's bare mass rather than the effective mass, and the bare charge (negative) rather than
the effective charge (positive) that enter into Eq. (1), is unexplained within the conventional theory of superconductivity.
In particular it implies that the superfluid carriers `undress' from their interaction with the ionic lattice\cite{undress,undress2}. Instead, the conventional theory asserts that the carriers are tightly coupled to the
lattice since the origin of the interaction that leads to superconductivity is precisely the interaction between the electrons and the ionic lattice.

Furthermore, for the magnetic field to develop when a rotating normal metal is cooled into the superconducting state, the superfluid electrons near the surface need to $slow$ $down$ in order to create the surface
current that gives rise to the magnetic field Eq. (1), $and$, negative charge needs to move $inward$ to satisfy mechanical equilibrium\cite{lorentz}.  The conventional theory does not explain
the origin of the forces giving rise to these effects, characterized as `quite absurd from the viewpoint of the perfect conductor concept' by Fritz London\cite{london}.

   \subsection{9. Inability to explain the Meissner effect}
   
   The Meissner effect is the most fundamental property of superconductors. When a superconductor is cooled in the presence of a static magnetic field,
   a spontaneous electric current near the surface of the superconductor develops that  nullifies the magnetic field in its interior\cite{meissner}. The literature on the
   conventional theory of superconductivity does not ever address
   nor answer the 
   following questions: (i) How do electrons near the surface of the sample acquire
   the superfluid velocity needed to screen the magnetic field in the interior?  (ii) How is angular momentum conserved in the process?
   These are fundamental questions that relate  to the very essence of the phenomenon of superconductivity.
   
   To the first question, a conventional superconductivity theorist may answer that because the final state with supercurrent flowing has lower free energy
  than  the initial state, the system will somewhow get there. However the supercurrent is a macroscopic effect and it should be possible to identify a macroscopic
   $force$ that leads electrons near the surface
   to start  moving  all in the same direction to give rise to the required current. There isn't such a force in the conventional theory of
   superconductivity. Concerning the second question,
    because the supercurrent in the final state carries mechanical angular momentum, and because the total angular momentum in the 
   normal state is zero, there exists a `missing angular momentum'\cite{missing}.  A conventional superconductivity theorist  may answer that the ionic lattice
   takes up the missing angular momentum. However the conventional theory offers no mechanism by which such an angular momentum transfer
   between superfluid electrons and the ionic lattice would take place.
   
Since 2003 I have pointed out repeatedly this inconsistency in the conventional theory\cite{lorentz,missing,m2,m3,m4,m5},
 to no avail. No answers to   these questions have been put forth by any of the believers in the 
   conventional theory of superconductivity.

  \subsection{10. Deviation from Occam's razor}
  
  Occam's razor is the philosophical principle that states that the explanation of any phenomenon should make as few assumptions as possible.
  Alternatively, that  the simplest solution to a problem is preferable to more complicated solutions. 
  However, as reviewed above, to explain all superconductors known today one needs many different mechanisms and 
  fundamentally different physical assumptions.

  Why is this implausible? Because there are fundamental characteristics of superconductors that $are$ shared by all of them, namely: the Meissner effect, 
  the Tao effect, the London moment,
  and the existence of macroscopic phase coherence (Josephson effect). These characteristics are remarkable and qualitatively different from the properties of 
  non-superconducting matter. It would be remarkable if nature had chosen to achieve these  properties in materials through many different 
  physical mechanisms
  and qualitatively different superconducting states.  The progress of science has shown again and again
  that true scientific advances in understanding always simplify previously existing   theories and unify the description of seemingly different phenomena.

  We can make a parallel here with atomic physics. The spectra of atoms is very complicated and certainly cannot be explained by a simple Balmer-like formula that works for 
  hydrogen only. However we don't need a different `mechanism' or theory to explain the atomic spectra of alkali metals, transition metals, rare gases, etc. All can be understood 
  from the same fundamental principles that were first understood in the context of the simplest atom, hydrogen.     Where is the `hydrogen atom' of superconductivity?
 
 \section{An alternative to BCS}
    
    At various points in this paper I have mentioned the theory of hole superconductivity\cite{mytheory,holesc1,undress,missing}.      Essential aspects of the theory are:
    
    (1) It applies  to all  superconducting materials.
    
    (2) Electron-hole asymmetry is the key to superconductivity; hole carriers in the normal state are necessary for superconductivity.
    
    (3) Electron-phonon interaction does not cause superconductivity; pairing is driven by a purely electronic mechanism associated with kinetic energy lowering\cite{kinetic}.
    
    (4) Material characteristics favorable for high $T_c$ are: (i) transport in the normal state dominated by hole carriers; (ii) excess negative charge in the substructures
    (e.g. planes) where conduction occurs\cite{molecule}.
    
   (5) The gap function versus energy has a slope of universal sign, giving rise to asymmetry in tunneling experiments of universal sign\cite{asymmetry}.
    
    (6) Superconductors expel negative charge from their interior towards the surface in the transition to superconductivity\cite{chargeexp}.
    
    (7) London electrodynamic equations are modified\cite{m4,electrodyn}. Macroscopic charge inhomogeneity and a macroscopic outward pointing electric field exist  in the 
    interior of superconductors. Applied electric fields are screened by the superfluid over a London penetration depth distance $\lambda_L$ rather than over the much shorter Thomas Fermi distance.  
    
    (8) A macroscopic spin current flows within a London penetration depth of the surface of superconductors, a kind of `zero point motion' of the superfluid\cite{m3}.
    
    (9) The spin-orbit interaction plays a fundamental role in superconductivity\cite{m5}.
    
    (10) Superfluid holes reside in mesoscopic orbits of radius $2\lambda_L$ and carry orbital angular momentum $\hbar/2$\cite{m3,lastpaper}.
    
     The theory offers transparent explanations for the Meissner effect\cite{missing},   the Tao effect\cite{taoyo}, the puzzles of rotating superconductors\cite{lorentz,spinc}, Chapnik's rule\cite{holeelec},
     and the variation of $T_c$ along the elements in the transition metal series\cite{transition1,transition2}.
     The `soft phonon' story\cite{softph} and the propensity of superconductors to be close to lattice instabilities\cite{instabilities}, conventionally understood as arising from strong electron-phonon interactions,
     are more simply explained from the fact that superconductors have nearly full bands and hence  a lot of electrons in $antibonding$ states\cite{bondcharge}. The same principle explains qualitatively
     why superconductivity is favored at high pressures: the externally applied pressure counters the outward pressure exerted by electrons occupying antibonding states, which would otherwise
     render the system unstable. As Bernd Matthias famously said\cite{instabilities}, ``From now on, I shall look for systems that should exist, but won't - unless one
can persuade them.''     The criteria given in  (4) above provide  guidelines in the search for new superconducting compounds, they explain why high $T_c$ is found in the cuprates and predict that high $T_c$ will be found
     in $MgB_2$ and $Fe-As$ compounds. They also predict\cite{dynh} that high $T_c$ will $not$ be found in $Li_{1-x}BC$\cite{predlibc,prediclibcprime,predlibc2} because it has
     far less negative charge in the planes than $MgB_2$.
        
  Examples of   experiments that could provide key evidence in support of this theory and against 
  conventional BCS theory are:  
    
  (1) Detection of spontaneous macroscopic electrostatic fields in or around superconductors, of magnitude comparable to the magnetic critical field ($H_c$ or $H_{c1}$) in cgs units.
  
  (2) Measurement of a macroscopic spin current in the ground state of a superconductor, of the predicted magnitude, namely  carrier density the superfluid density and
  carrier speed given by the speed of carriers in the critical charge current of the superconductor.
  
  (3) Measurement of a much steeper plasmon dispersion relation in the superconducting state than in the normal state\cite{electrodyn}.
  
  (4) Detection of ionizing radiation emitted by a superconductor of large volume under non-equilibrium conditions, of frequencies up to $\omega=  0.511 MeV/ \hbar$\cite{ionizing}.
  
  As a historical footnote I point out that several elements of this theory were part of or are related to pre-BCS proposed explanations of
  superconductivity, namely: (i) Heisenberg\cite{heisenberg} and others proposed that currents exist in the ground state of superconductors, albeit charge rather than spin
  currents; (ii) Born and Cheng\cite{born} proposed that superconductivity could only occur when the Fermi surface is close to the edges of the
  Brillouin zone; (iii) Slater\cite{slater} proposed that electrons in superconductors reside in orbits of radius $\sim 137$ lattice spacings.

  \section{Discussion}

  This paper focused on  BCS theory, however it is clear that more generally it may apply to all  areas of contemporary science,
i.e.  that the same factors at play in the Madoff case may be allowing for the preservation and growth of many flawed scientific theories {\it at the present time}. With the growth and 
  specialization of knowledge, incoming   students 
  have to increasingly rely on previously established scientific results as `gospel',
  and they increasingly have to rely on `gatekeepers' (professors, mentors, established scientists) 
  to guide them into the world of science. The  gatekeepers have a vested
  interest in preserving the status quo. A beginning scientist with a revolutionary idea that could prove many established scientists wrong is likely to be strongly discouraged from pursuing it, and if s/he persisted would simply be  denied 
  entrance to the profession by being unable to secure a job. By the time a scientist is `established' he or she has usually been sufficiently conditioned  to conform to the 
  established truths. 
    
 For the case of BCS, to facilitate a faster and softer landing, I suggest  that: (1) Journal editors   should look more favorably than they have up to now
  at papers suggesting inadequacies of BCS theory, and keep in mind the vested interests of referees that are likely to write negative
  reports on such papers.  To the extent that such papers can be published in mainstream
  publications, they will encourage  physicists,   the younger generation as well as some of the long-time experts that may have started having doubts about
  BCS in view of the recent experimental discoveries, to consider alternatives to the conventional BCS theory.
  (2) Grant allocation officers  should consider funding both experimental and theoretical research work that calls into question the conventional BCS theory.
  Currently experimentalists are reluctant to devote resources to experiments that don't conform to the conventional theory assumptions.
   (3) Conference organizers   should 
   consider inviting speakers whose research questions the validity of BCS theory for conventional superconductors rather than shun such topics.

 The half-century old BCS theory has proven incapable of ever predicting a high temperature superconductor. It offers no useful guidelines in the search for new superconducting compounds.
 It has proven incapable of explaining the superconductivity of ten families of compounds discovered in the last thirty years. It can't explain the Meissner effect nor the Tao effect nor Chapnik's rule
 nor rotating superconductors. The field of superconductivity is in crisis\cite{kuhn}.
 It is high  time to consider the possibility that the lack of progress in understanding high $T_c$ cuprates and other `unconventional' superconductors may be due to the fact that
 `conventional' superconductors are not understood either. It is high time to seriously consider the possibility that BCS theory provides no real understanding of 
 the superconductivity of `conventional' materials because it is fundamentally flawed,  and that it may be destined to be overhauled 
 just as other established scientific theories of the past have been overhauled.


\begin{references}
\bibitem{silence} It has been impossible
              for me during  the last twenty years  to publish any paper  in those journals,
              not even a reader's letter in the case of Nature and Science,  that would question the validity of BCS 
              applied to the conventional superconductors, despite repeated attempts to do so. 
              Such papers I 
              sometimes manage to publish in less `prestigious' journals  and they are  ignored by the mainstream physics community.
\end{references}
\end{document}